%% file: zapaper.tex
\documentclass[a4paper]{JHEP3}
\usepackage{amsmath,amssymb}
\usepackage{epsfig,cite}

   \let\b=\beta

\newcommand{\be}{\begin{equation}}
\newcommand{\ee}{\end{equation}}
\newcommand{\bea}{\begin{eqnarray}}
\newcommand{\eea}{\end{eqnarray}}
\newcommand{\ba}{\begin{array}}
\newcommand{\ea}{\end{array}}

\def\fm{\,{\rm fm}}

\def\rmO{\textrm{O}}

\newcommand{\cA}{c_{\rm A}}

\newcommand{\csw}{c_{\rm sw}}
\newcommand{\za}{Z_{\rm A}}

\newcommand{\fig}[1]{Fig.~\ref{#1}}
\newcommand{\tab}[1]{Table~\ref{#1}}
\newcommand{\sect}[1]{Section~\ref{#1}}

\newcommand{\ar}{A_{\rm R}}
\newcommand{\ai}{A_{\rm I}}
\newcommand{\mq}{m_{\rm q}}
\newcommand{\funci}[1]{f_{\rm #1}^{\rm I}}
\newcommand{\tfunci}[1]{\tilde f_{\rm #1}^{\rm I}}
\newcommand{\func}[1]{f_{\rm #1}}
\newcommand{\fv}{f_{\rm V}}

\def\rmd{{\rm d}}
\def\rmO{{\rm O}}

\def\psibar{\overline{\psi}}

\def\zetabar{\bar{\zeta}}
\def\zetaprime{\zeta\kern1pt'}
\def\zetabarprime{\zetabar\kern1pt'}

\def\dirac#1{\gamma_{#1}}

\def\ba{b_{\rm A}}
\def\cA{c_{\rm A}}
\def\csw{c_{\rm sw}}

\def\za{Z_{\rm A}}

\def\zv{Z_{\rm V}}

\def\bx{{\bf x}}
\def\by{{\bf y}}

\def\bu{{\bf u}}
\def\bv{{\bf v}}

\def\O{\mathcal O}
\def\R{\mathcal R}

\def\half{{\textstyle\frac12}}\def\op{\O}
\def\oprime{\O'}

\def\ts{\textstyle}
\def\drvtilde#1#2{{\tilde{\partial}_{#1}^{#2}}}

\title{Non--perturbative renormalization of the axial current
with dynamical Wilson fermions}

\author{Michele Della Morte$\,^a$, Roland Hoffmann$\,^a$, Francesco Knechtli$\,^a$,}
\author{\vspace*{-6.5mm}Rainer Sommer$\,^b$ and Ulli Wolff$\,^a$\\
$^a\,$Institut f\"ur Physik, Humboldt Universit\"at,
Newtonstr. 15, 12489 Berlin, Germany\\
$^b\,$DESY,
Platanenallee 6, 15738 Zeuthen, Germany\\
\vspace*{10mm}
\center \epsfig{file=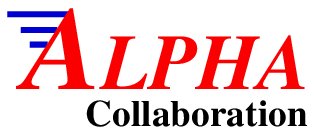,width=29mm}\\[4mm]
}

\preprint{HU-EP-05/23\\ SFB/CPP-05-18\\ DESY 05-064}

\abstract{
We present a new normalization condition for the axial current,
derived from the PCAC relation with non--vanishing quark
mass. This condition is expected to reduce mass effects in
the chiral extrapolation of the results for the normalization factor
$\za$. 
The application to the two--flavor theory with improved Wilson
fermions shows that this expectation is indeed fulfilled.
Using the Schr\"odinger functional setup we calculate $\za(g_0^2)$
as well as the vector current normalization factor $\zv(g_0^2)$
for $\b=6/g_0^2 \geq 5.2$.
}


\begin{document}

\input{sect1.tex}
\input{sect2.tex}
\input{sect3.tex}

\input{sect4.tex}

\bigskip

\acknowledgments

We are grateful to Martin L\"uscher and Stefan Sint for illuminating discussions
and Hartmut Wittig for communicating details about the simulations
performed for \cite{Luscher:1996jn}.
We thank NIC/DESY Zeuthen for allocating computer time on the
APEmille machines for this project.
This work is part of the ALPHA collaboration's research
programme. It was supported by the Deutsche Forschungsgemeinschaft
in the
form of the Graduiertenkolleg
GK271 and the SFB/TR 09-03.  All the computer runs were carried
out on
machines of the APEmille series at DESY Zeuthen. We thank the staff at
DESY Zeuthen for their help.

\appendix
\input{disco}
\section{List of simulation parameters and results}
\label{simdata}
In \tab{longtable} the simulation results for $\za$ and $\zv$
are collected. The number of measurements $N_{\rm meas}$
explicitly contains the number of replica
and  $\tau_{\rm meas}$ is the number of (unit length) trajectories
between consecutive measurements.

\input{table}

\bibliography{../refs}           
\bibliographystyle{JHEP}   

\end{document}

%% file: sect1.tex
\section{Introduction}

In recent years Wilson's formulation of lattice QCD \cite{Wilson:1974sk}
has matured to a stage where simulations with
light dynamical fermions
are within reach \cite{AliKhan:2001tx,Allton:2001sk,Aoki:2002uc}.
For light quarks some of the most interesting physics at low energies is
associated with the pseudo--scalar sector of the theory, whose dynamics
is governed by chiral symmetry.
It is therefore important to study how and to which extent this symmetry
can be realized in our chosen regularization.

For Wilson fermions the (local) isovector axial current is not
associated with a symmetry of the lattice action and consequently
it does not satisfy
continuum Ward-Takahashi identities.
However, the latter can be restored up to cutoff--effects
(i.e. powers of the lattice spacing)
through a finite rescaling of the axial current
\cite{Bochicchio:1985xa}.
Its normalization factor $\za(g_0^2)$ is exactly obtained
in this way
by enforcing one particular continuum Ward identity at finite lattice
spacing.

Although a perturbative estimate of $\za$ is available, most
of today's simulations are performed at bare gauge couplings
$g_0^2\simeq1$, where bare perturbation theory cannot be expected
to work. To keep systematic effects in physical observables
under control, it is therefore mandatory
to determine $\za$ non--perturbatively.

The approach used by the ALPHA Collaboration in the quenched
approximation \cite{Luscher:1996jn}
can obviously also be applied to the dynamical case.
Here we improve this method by deriving a normalization condition
at finite quark mass, which provides a technical advantage
when extrapolating the results to the chiral limit.
Using the Schr\"odinger functional setup we compute $\za(g_0^2)$
for the relevant range of bare gauge
couplings. In addition, we check the matching of our non--perturbative
estimate with 1--loop perturbation theory through
simulations at large $\beta$. Systematic effects and in
particular $\rmO(a^2)$ uncertainties arising from a variation
of the normalization condition are also considered.
These turn out to be rather large for $\beta\lesssim5.4$.

The paper is organized as follows.
After recalling the axial Ward identities we review the
normalization condition used in \cite{Luscher:1996jn} before
introducing the new ''massive'' condition.
Numerical results for $\za$ and $\zv$ are presented in \sect{numerics},
where we also provide interpolating formulae.
In \sect{conclusions} our conclusions are summarized and
we discuss possible applications of the results.

%% file: sect2.tex
\section{Theory}

We consider an isospin doublet of quarks with mass $m$ 
and proceed formally in the continuum theory. The notation employed
follows that of
\cite{Luscher:1996sc} and a pedagogical introduction can be found
in~\cite{Luscher:1998pe}. 

\subsection{The axial Ward identity in the continuum}

By performing local infinitesimal
transformations of the quark and anti--quark fields in the Euclidean
functional integral one derives the Ward identities associated with
the flavor chiral symmetry of the action. An axial transformation
gives the partially conserved axial current (PCAC) relation
\begin{eqnarray}
\langle\partial_\mu A_\mu^a(x)\O\rangle&=&2m\langle P^a(x)\O\rangle,
\quad\textrm{where}\label{PCAC}\\
A_\mu^a(x)=\psibar(x)\dirac\mu\dirac5\half\tau^a\psi(x)\;,&&
P^a(x)=\psibar(x)\dirac5\half\tau^a\psi(x)
\end{eqnarray}
are the axial current and the pseudoscalar density, respectively. Here
$\tau^a$ are the Pauli matrices acting on the flavor indices of
the quark fields. The relation (\ref{PCAC}) holds for any operator
$\O$ built from the basic fields in a region not containing the
point $x$.

Let $\R$ be a space--time region with smooth boundary $\partial
\R$ where the symmetry transformations are applied. The axial
current $A_\nu^b(y)$ is inserted as an internal operator
($y\!\in\!\R$). If $\O_{\rm ext}$ denotes a polynomial in the
basic fields outside this region the integrated form of the axial
Ward identity is
\begin{eqnarray}
\int_{\partial \R}\!\!\!\textstyle\rmd\sigma_\mu(x)\,\Big\langle
A_\mu^a(x)A_\nu^b(y)\O_{\rm ext}
\Big\rangle\qquad\qquad\nonumber\\
\qquad\qquad-2m\int_R\!\!\rmd^4x\, \Big\langle
P^a(x)A_\nu^b(y)\O_{\rm ext}\Big\rangle&=&
i\epsilon^{abd}\Big\langle V_\nu^d(y)\O_{\rm
ext}\Big\rangle\;.\label{WI}
\end{eqnarray}
The right--hand side of (\ref{WI}) originates from the variation of
the internal operator $A_\nu^b(y)$ since under chiral
transformation the axial current linearly combines with the vector current
\begin{equation}
V_\nu^d(y)=\psibar(y)\dirac\nu\half\tau^d\psi(y)\;.
\label{localV}
\end{equation}
In the relation (\ref{WI}) we set $\nu=0$
and choose the region $\R$ to be the space--time volume
between the hyper--planes at $y_0\!-\!t$ and $y_0\!+\!t$.
After introducing a spatial integration over $\by$ and contracting the
isospin indices with the totally antisymmetric tensor $\epsilon^{abc}$
we arrive at
\begin{eqnarray}
\int\!\rmd^3\by\int_{\partial
\R}\!\!\!\!\textstyle\rmd\sigma_\mu(x)\,\epsilon^{abc}\Big\langle
A_\mu^a(x)A_0^b(y)\O_{\rm ext}
\Big\rangle\qquad\qquad\nonumber\\
\qquad\quad-2m\int\!\rmd^3\by\int_\R\!\!\rmd^4x\,
\epsilon^{abc}\Big\langle P^a(x)A_0^b(y)\O_{\rm
ext}\Big\rangle&\!\!=\!\!& 2i\!\int\!\rmd^3\by\Big\langle V_0^c(y)\O_{\rm
ext}\Big\rangle\;.\qquad\label{WI2}
\end{eqnarray}
Note that the integral over $\R$ includes a contact term at the
point $x=y$. However, power counting and the operator product
expansion tell us that the correlation function multiplying
the mass has no non--integrable short--distance
singularity.

The above equation can be simplified by combining the two
contributions from the surface integral over $\partial\R$.
We first notice that with periodic boundary conditions in
the spatial directions
$\int\!\rmd^3\bx\,\partial_kf(x)=0$ for any $f(x)$
and thus the integrated form of the PCAC relation (\ref{PCAC}),
now written as an operator identity, reads
\begin{eqnarray}
      \int\!\!\rmd^3\bx\, \partial_0 A_0^a(x)=
      \int\!\!\rmd^3\bx\, \partial_\mu A_\mu^a(x)=
2m\!\!\int\!\!\rmd^3\bx\, P^a(x)\;.
\end{eqnarray}
Integrating this relation from $y_0\!-\!t$ to $y_0$
then results in
\begin{eqnarray}
\int\!\!\rmd^3\bx\, A_0^a(y_0\!-\!t,\bx)&\!\!=\!\!&
\int\!\!\rmd^3\bx\, A_0^a(y_0,\bx)-2m\int_{y_0-t}^{y_0}\!\!\rmd x_0
\int\!\!\rmd^3\bx\, P^a(x_0,\bx)\label{shift}\;.
\end{eqnarray}
With the region $\R$ defined as above, the integration over the
lower surface
in (\ref{WI2}) involves current insertions at $y_0\!-\!t$ and $y_0$,
which can be shifted to $y_0$ and $y_0\!+\!t$
by using (twice) the partial conservation of the axial current
in the form of
(\ref{shift}).
The final result is
\begin{eqnarray}
\int\!\!\rmd^3\by\,\rmd^3\bx\, \epsilon^{abc}
\Big\langle A_0^a(y_0\!+\!t,\bx)A_0^b(y)\O_{\rm ext}\Big\rangle
\qquad\qquad\nonumber\\
-2m\!\int\!\!\rmd^3\by\,\rmd^3\bx\!\int_{y_0}^{y_0+t}\!\!\!\!\!\rmd x_0\,
\epsilon^{abc}\Big\langle P^a(x)A_0^b(y)\O_{\rm ext}\Big\rangle
&\!=\!&i\displaystyle\!\int\!\!\rmd^3\by\
\Big\langle V_0^c(y)\O_{\rm ext}\Big\rangle\;.\qquad\label{WI3}
\end{eqnarray}
Note that this formal manipulation is not invalidated by the contact
term in (\ref{WI2}), which now appears at the lower integration limit.

\subsection{Normalization of the axial current on the lattice}

A normalization condition for the axial current on the lattice is
derived by demanding that eq.~(\ref{WI3}) in terms of the renormalized
currents holds at non--zero lattice spacing. While this assumes the
knowledge of the renormalization factor of the local vector current, it will
become clear that with our choice of $\op_{\rm ext}$ the latter can be
easily computed.
In the improved theory, eq.~(\ref{WI3}) then defines the normalization
factor up to $\rmO(a^2)$
uncertainties.
We construct the relevant matrix elements in the framework of the
Schr\"odinger functional \cite{Luscher:1992an,Sint:1993un}.
The notation is taken over from \cite{Luscher:1996sc}, to which we
refer for any unexplained symbol.

Our starting point is the improved axial current
\begin{equation}
\label{imprcurrent}
(\ai)_\mu^a=A_\mu^a+a \cA\drvtilde\mu{} P^a\;,
\end{equation}
where $\drvtilde\mu{}$ denotes the symmetric lattice derivative.
In a
mass--independent renormalization scheme
the renormalized (improved) current takes the
form~\cite{Jansen:1995ck,Luscher:1996sc}
\begin{equation}
(\ar)_\mu^a=\za(1+b_{\rm A} a \mq)(\ai)_\mu^a\;.\label{rencurrent}
\end{equation}
Here $\mq=m_0-m_c$
is the bare subtracted quark mass, which on the lattice we need to
distinguish from the current quark mass $m$, derived from a
discretized version of (\ref{PCAC}).

In implementing (\ref{WI3}) on the lattice, we choose the isovector scalar
\begin{equation}
\label{oext}
\O^c_{\rm ext}=-\frac1{6L^6}\epsilon^{cde}\oprime^d\O^e
\end{equation}
as external operator. It is built from
the zero momentum sources
\begin{eqnarray}
\op^a&=&\ts a^6\sum_{\bf u,v}
\zetabar(\bu)\dirac{5}\frac{1}{2}
\tau^a\zeta(\bv)\quad \textrm{and}\nonumber\\
\oprime^a&=&\ts a^6\sum_{\bf u,v}
\zetabarprime(\bu)\dirac{5}\frac{1}{2}
\tau^a\zetaprime(\bv),
\end{eqnarray}
where $\zeta$ and $\zetaprime$ are the quark fields at the SF boundary
$x_0=0$ and $x_0=T$, respectively. The isospin index of the external
operator is contracted with the free index in (\ref{WI3}). As
was shown in \cite{Luscher:1996jn}, isospin symmetry in the form of a
vector Ward identity implies that with this
setup the right-hand side of
eq.~(\ref{WI3}) simplifies to the boundary-to-boundary correlation
function
\begin{equation}
f_1=-\frac1{3L^6}\langle\oprime^a\op^a\rangle\label{f1}
\end{equation}
up to $\rmO(a^2)$. This gives the vector current normalization condition,
which we will discuss in more detail in the next section.

For the normalization condition
derived in \cite{Luscher:1996jn} (''old'' condition) the mass $\mq$
was set to zero in
(\ref{WI3}) and (\ref{rencurrent}). Since the normalization of the boundary
fields cancels as they appear on both sides of (\ref{WI3}), with
the external operator (\ref{oext}) the old normalization
condition can be written as 
\begin{eqnarray}
\za^2\funci{AA}(y_0\!+\!t,y_0)=f_1+\rmO(a^2),\label{normcon1}
\end{eqnarray}
with the correlation function
\begin{eqnarray}
\funci{AA}(x_0,y_0)&=&-\frac{a^6}{6L^6}\sum_{\bx,\by}
\epsilon^{abc}\epsilon^{cde}
\Big\langle \oprime^d (\ai)_0^a(x)(\ai)_0^b(y)\O^e\Big\rangle\;.
\label{fAAI}
\end{eqnarray}
In \cite{Luscher:1996jn} the choice $y_0\!=\!t\!=\!T/3$ was made
in order to maximize the distance between the current insertions
and thus reduce lattice artifacts.
Since the mass term in the Ward identity is neglected entirely,
an evaluation of this normalization condition at non--vanishing quark mass
leads to errors of $\rmO(r_0 m)$ in addition to the usual
$\rmO(a^2)$ errors. Here we have used $r_0$, introduced in \cite{Sommer:1993ce},
as a typical low--energy scale.

\subsection{New normalization condition}
\label{sectnew}

We now repeat the above steps using the full axial Ward identity to derive
a normalization condition for the axial current. The additional term in
eq.~(\ref{WI3}) proportional to the quark mass
results in a new correlation function $\tfunci{PA}(y_0\!+\!t,y_0)$, which is similar to
(\ref{fAAI}) but involves a temporal sum
\begin{eqnarray}
\tfunci{PA}(y_0\!+\!t,y_0)&=&-\frac{a^7}{6L^6}\! \sum_{x_0=y_0}^{y_0+t}
\!w(x_0)\!
\,\sum_{\bx,\by}\epsilon^{abc}\epsilon^{cde}\Big\langle
\oprime^dP^a(x)(\ai)_0^b(y)\O^e\Big\rangle\label{ftilde}\;,\\[2mm]
\textrm{where } w(x_0)&=&\left\{
\begin{array}{r@{,\quad}l}
1/2 & x_0=y_0\ \textrm{ or }\ y_0\!+\!t\\
1       & \textrm{otherwise}\;,
\end{array}\right.
\end{eqnarray}
is introduced in order to implement the trapezoidal rule for discretizing
integrals.
Due to the aforementioned
contact--terms on-shell $\rmO(a)$ improvement does not apply
to the expression~(\ref{ftilde}). Therefore, although this new condition
removes the $\rmO(r_0m)$ uncertainties in the determination of $\za$ at
finite mass, other $\rmO(am)$ ambiguities appear.
In a quenched study \cite{Hoffmann:2003mm} this residual mass
dependence was shown to be very small.

It follows from the PCAC relation that the product $mP^a$
renormalizes with the same factor as the axial current and hence
the new massive normalization condition can be written as\\[-7mm]
\begin{eqnarray}
\za^2(1\!+\!\ba a\mq)^2\Big(\funci{AA}(y_0\!+\!t,y_0)
-2m\tfunci{PA}(y_0\!+\!t,y_0)\Big)\!\!&\label{normcon2}
=&f_1+\rmO(am)+\rmO(a^2)\;.\qquad
\end{eqnarray}
In our calculation we neglect the coefficient $\ba$, which is only known
perturbatively. This choice only changes the $\rmO(am)$ effects.
The correlation functions for this new normalization condition
can be evaluated with the same
geometry as for (\ref{fAAI}),
i.e. the volume integral covers the middle third of the
temporal extension of the lattice.

We now proceed to some
technical aspects concerning the
evaluation of $\funci{AA}$ and $\tfunci{PA}$.
Inserting
the expression (\ref{imprcurrent}) for the improved axial
current into the correlation functions (\ref{fAAI})
and (\ref{ftilde}) we have
\begin{eqnarray}
\funci{AA}(x_0,y_0)\!&=&\!\func{AA}(x_0,y_0)
+a\cA\Big[\drvtilde0x\,\func{PA}(x_0,y_0)\nonumber
+\drvtilde0y\,\func{AP}(x_0,y_0)\Big]\\
&&\!\!\phantom{\func{AA}(x_0,y_0)}+a^2\cA^2\,\drvtilde0x\,\drvtilde0y\,\func{PP}(x_0,y_0),
\label{fAAI_decompose}\\[2mm]
\tfunci{PA}(y_0\!+\!t,y_0)\!&=&\!a\!\!
\sum_{x_0=y_0}^{y_0+t}\!w(x_0)\Big[\func{PA}(x_0,y_0)
+a\cA\drvtilde0y\func{PP}(x_0,y_0)\Big],\\[1mm]
\textrm{where: }\func{XY}(x_0,y_0)\!&=&\!-\frac{a^6}{6L^6}
\sum_{\bf x,\bf y}
\epsilon^{abc}\epsilon^{cde}
\Big\langle\!\oprime^d
X^a(x)Y^b(y)
\O^e\!
\Big\rangle,\ X,Y\!\in\!\{A_0,P\}.\qquad\label{fXY}
\end{eqnarray}
Performing the Wick contractions for this correlation function, one finds that
as a consequence of the isospin structure of the boundary composite fields
$\op^a$ and $\oprime^a$ only
six quark diagrams contribute.
\EPSFIGURE[ht]{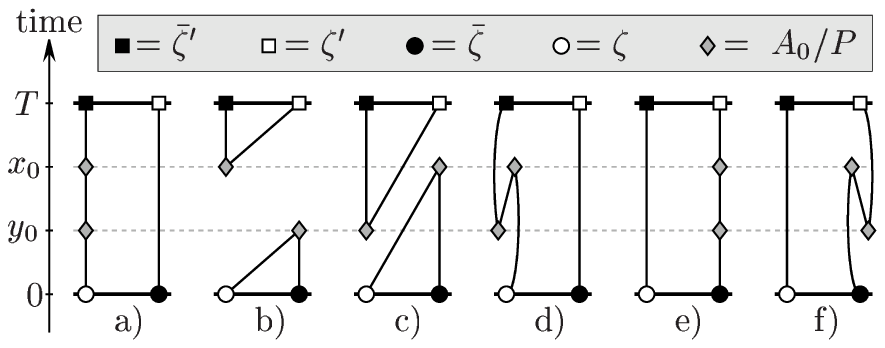,width=106mm}{
A graphical representation of the possible Wick
contractions for the correlation
functions $f_{XY}(x_0,y_0)$.
The gray diamonds indicate the insertions of $Y$ and $X$ at times
$y_0$ and $x_0$.\label{fig:contractions}}
Those are shown in Figure
\ref{fig:contractions}. Among them there are also two
disconnected diagrams $b)$ and $c)$, where no propagator
connects the $x_0=0$ and $x_0=T$ boundary fields.
Exploiting the conservation of the axial current and
making use of the operator product expansion, one can show
\cite{disco}
that in the continuum the diagrams b) and c) cancel for
vanishing quark mass (the detailed argument is
given in Appendix \ref{appA}).
This implies that on the lattice they give only an
$O(a^2)$ contribution to $\funci{AA}$ and $\tfunci{PA}$
in the improved theory.
An alternative definition of $\za$ is thus obtained by using
only the connected part of the correlation functions in
(\ref{normcon2}).

\subsection{The vector current}

Isospin symmetry, e.g. in the form of an
integrated vector Ward identity, implies that
on the lattice
\be
\zv(1+b_{\rm V}a\mq)\fv(x_0)=f_1+\rmO(a^2)\;,\label{veccond}
\ee
where $\fv(x_0)$ is the Schr\"odinger functional correlation
function
\be
\fv(x_0)=\frac{a^3}{6L^6}\sum_\bx i\epsilon^{abc}
\langle\oprime^a V_0^b(x)\op^c\rangle\;.\label{fv}
\ee
Note that with spatially periodic boundary conditions,
the term proportional to the improvement coefficient $c_{\rm V}$
\cite{Luscher:1996sc,Sint:1997jx}
does not contribute to (\ref{fv}).
In our implementation we also neglect $b_{\rm V}$ and extrapolate the simulation
results to zero mass to obtain $\zv$ from (\ref{veccond})
in the chiral limit. The insertion point $x_0$ in (\ref{veccond}) is taken
to be $T/2$. The procedure is analogous to the one
adopted in the quenched case \cite{Luscher:1996jn}. There the slope (in
$a\mq$)
of $f_1/(\zv\fv)$ gives an estimate of $b_{\rm V}$. This is no longer true
in the unquenched
theory as the improvement coefficient $b_{\rm g}$ is different from zero
\cite{Luscher:1996sc} and thus the slope in $a\mq$ is not due to
$b_{\rm V}$ alone.

%% file: sect3.tex
\section{Numerical Results}
\label{numerics}

We performed our simulations using non--pertur\-ba\-tively improved Wilson fermions
\cite{Luscher:1996ug,Luscher:1996sc,Jansen:1998mx,Yamada:2004ja} and
the plaquette gauge action.
The clover coefficient $\csw$ has been set
to the value from \cite{Jansen:1998mx,Yamada:2004ja}
and for the axial current improvement
constant $\cA$ we have used the recently determined
non--perturbative estimate \cite{DellaMorte:2005se}.
Adopting the same setup as in the quenched computation
\cite{Luscher:1996jn,Hoffmann:2003mm} we choose $T=9/4L$
with periodic boundary conditions
in space. The background field is set to zero.

Concerning the algorithm,
we employed the HMC with two pseudo--fermion fields as proposed in
\cite{Hasenbusch:2001ne} and studied in the Schr\"odinger functional
in \cite{DellaMorte:2003jj} as well as the PHMC algorithm
\cite{deForcrand:1996ck,Frezzotti:1997ym}.
This choice has been motivated in \cite{DellaMorte:2004hs},
where PHMC has been shown to be more efficient
at the coarsest lattice spacing, since there the spectrum of
the improved Wilson--Dirac operator is affected by large
cutoff effects in the form of unphysically small eigenvalues.

Summarizing the discussion from \sect{sectnew}, we choose the following non--perturbative
definitions of the isovector normalization factors
\bea
\za(g_0^2)&=&\lim_{m\rightarrow0}\sqrt{f_1}\left[\funci{AA}(2T/3,T/3)
  -2m\tfunci{PA}(2T/3,T/3)\right]^{-1/2}\;,\label{zadef}\\
\zv(g_0^2)&=&\lim_{m\rightarrow0}\frac{f_1}{\fv(T/2)}\;.\label{zvdef}
\eea
The quark mass $m$ is defined as in \cite{Luscher:1996sc} and we average
it over three time--slices
\footnote{Four time--slices for odd $T$.} around $T/2$.
The same averaging is used to reduce the statistical error of  $\zv$.
Note that in the case of $\za$ a similar time average would require
additional components of the quark propagator. According to the
discussion at the end of section {\ref{sectnew} we define another
normalization factor
$\za^{\,\rm con}$ through (\ref{zadef}), where
we now drop the disconnected quark diagrams
in the correlation functions.

As discussed in \cite{Luscher:1996jn}, 
we need to evaluate the normalization conditions
on a line of constant physics, keeping all
length scales fixed.
This ensures
that the $\rmO(a^2)$ ambiguities in the normalization
factors vanish smoothly when the perturbative regime is
approached.
In addition, the normalization conditions have to be
set up at zero quark mass since we are aiming
for a mass--independent renormalization scheme.

To maintain a line of constant physics we need to know
the three values of $g_0^2$ that yield constant $L$
for $L/a=8,\,12,\,16$. Our value of $L$ is determined by starting
with the popular coupling
$\beta=6/g_0^2=5.2$ at $L/a=8$. The readjustments of the bare coupling
needed to achieve lattice spacings smaller by factors of $8/12$ and $8/16$
we take from the 3--loop perturbative formula \cite{Bode:2001uz}
\begin{eqnarray} \label{e:L3l}
\frac{a(g_0^2)}{a((g_0')^2)}&=&e^{ -[g_0^{-2}-(g_0')^{-2}]/ 2b_0}
[g_0^2/(g_0')^2]^{-{b_1/2b_0^2}}\,
\left[\, 1 + q\, [g_0^2 -(g_0')^2] + {\rm O}\left((g_0')^4\right)\,\right], 
\\
&& q = 0.4529(1)\,,\quad g_0<g_0'\;. \nonumber
\end{eqnarray}
For the range of bare couplings involved
this is also consistent with the non--perturbative coupling dependence
of $r_0/a$ as checked in \cite{DellaMorte:2005se}. It should in any case be clear to the
reader
that small deviations from the constant physics line here
influence the quality of improvement (size of remaining $a^2$-effects)
but do not imply any systematic errors of the continuum results.

In addition to
three matched lattice sizes $L/a=8,\,12,\,16$ at $\beta=5.2,\,
5.5,\, 5.715$, we simulated at
three larger values of $\beta$ and fixed $L/a=8$, which
corresponds to very small volumes.
This was done in
order to verify that our non--perturbative estimate
smoothly connects to the perturbative predictions
\cite{Gabrielli:1990us,
Gockeler:1996gu,sint:notes}
\bea
\za&=&1-0.116458\,g_0^2+\rmO(g_0^4)\;,\label{pta}\\
\zv&=&1-0.129430\,g_0^2+\rmO(g_0^4)\;.\label{ptv}
\eea
Our simulation
parameters as well as the final results for $\za$ and $\zv$
are collected in \tab{t_simpar}, where we also include
data from simulations at $\beta=5.29$ and slightly mismatched
volume. The latter are only used to qualitatively
confirm the observed rapid
change of $\za(g_0^2)$ in this region of the coupling.
\TABULAR[t]{|c|rrlll|}{
    \hline
$\beta$ &
\multicolumn{1}{c}{$L/a$} &
\multicolumn{1}{c}{$T/a$} &
\multicolumn{1}{c}{$\kappa_c\qquad$} &
\multicolumn{1}{c}{$\za\qquad$} &
\multicolumn{1}{c|}{$\zv\qquad$} \\
    \hline
5.200 &  8 & 18& 0.135856(18)& 0.7141(123)    & 0.7397(12)      \\
5.500 & 12 & 27& 0.136733(8) & 0.7882(35)(39) & 0.7645(22)(18)  \\
5.715 & 16 & 36& 0.136688(11)& 0.8037(38)(7)  & 0.7801(15)(27)  \\\hline
5.290 &  8 & 18& 0.136310(22)& 0.7532(79)     & 0.7501(13)    \\\hline
7.200 &  8 & 18& 0.134220(21)& 0.8702(16)(7)  & 0.8563(5)(45)   \\
8.400 &  8 & 18& 0.132584(7) & 0.8991(25)(7)  & 0.8838(13)(45)  \\
9.600 &  8 & 18& 0.131405(3) & 0.9132(11)(7)  & 0.9038(3)(45)   \\\hline
}
{Results for the chiral extrapolations of $\za$ (\ref{zadef})
and $\zv$ (\ref{zvdef}) and estimates for the critical hopping
parameter $\kappa_c$.\label{t_simpar}}
The results in \tab{t_simpar} are obtained through an interpolation
or slight
extrapolation in the quark mass. The first error we quote for $\za$
and $\zv$ is statistical and the second represents our estimate
of the systematic error, which originates from deviations from the constant physics
condition. It is discussed in the next section.
The complete set of simulations results and parameters is given
in \tab{longtable} in appendix \ref{simdata}.

As expected from
the above arguments and already verified in the quenched
case \cite{Hoffmann:2003mm}, for the new normalization
condition the data exhibit very little dependence on the
quark
mass. Consequently, uncertainties in the location of
the critical point do not propagate to the determination
of $\za$.
That this does not hold for the previously used
normalization condition \cite{Luscher:1996jn} can
be inferred from \fig{fig:extrapol}.
\EPSFIGURE[!h]{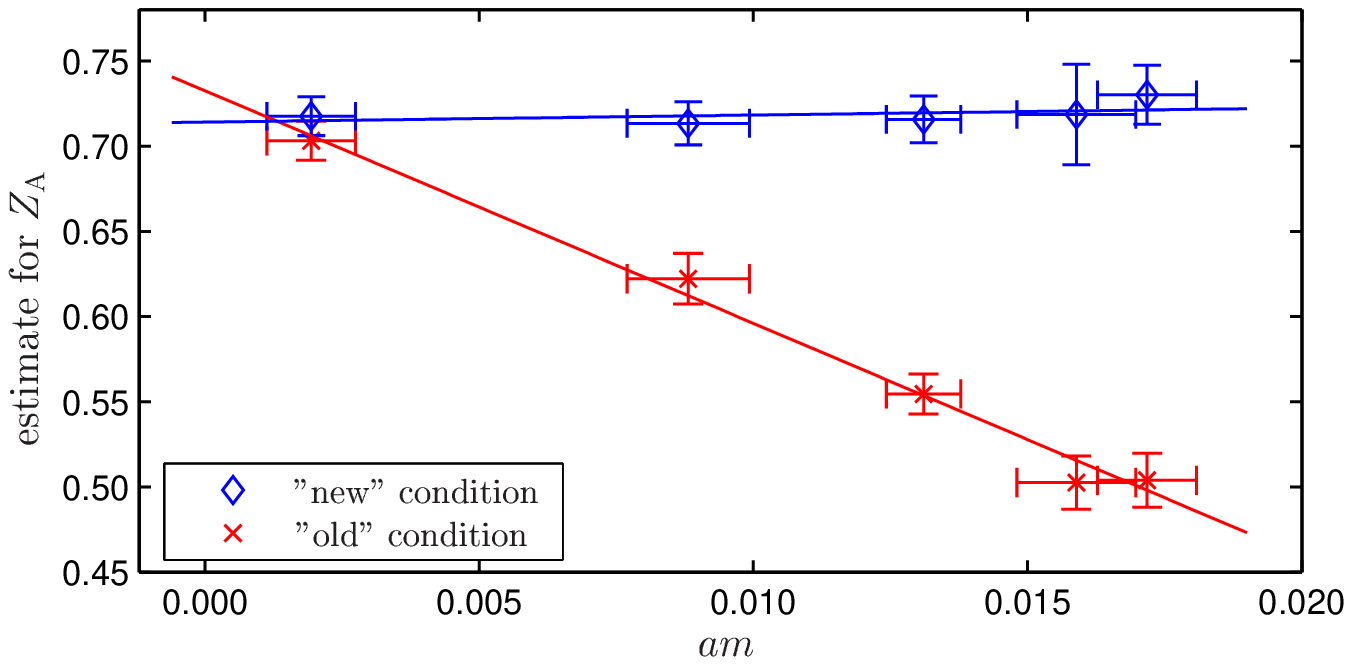,width=130mm}{
Comparison of the chiral extrapolation at $\beta=5.2$
using the new
and old normalization conditions for $\za$.
\label{fig:extrapol}}
There we show the chiral
extrapolation at $\beta=5.2$, where we had to use
the PHMC algorithm in order to obtain reliable
error estimates. With the HMC
algorithm the error analysis would have been tainted by
the rare occurrence of very small eigenvalues
\cite{DellaMorte:2004hs}.
While for the new normalization condition
the slope in $am$ is consistent
with zero,
the estimate of $\za$ from the old condition changes
by $30\%$ in the (small) mass range shown.
We anyway see that for $am\lesssim0.02$ all
mass effects show a linear behavior.
For $\beta=5.5$ the extrapolation is similar to
the one shown and at all other gauge couplings
we can in fact interpolate from two simulations
very close to the critical point.

\EPSFIGURE[ht]{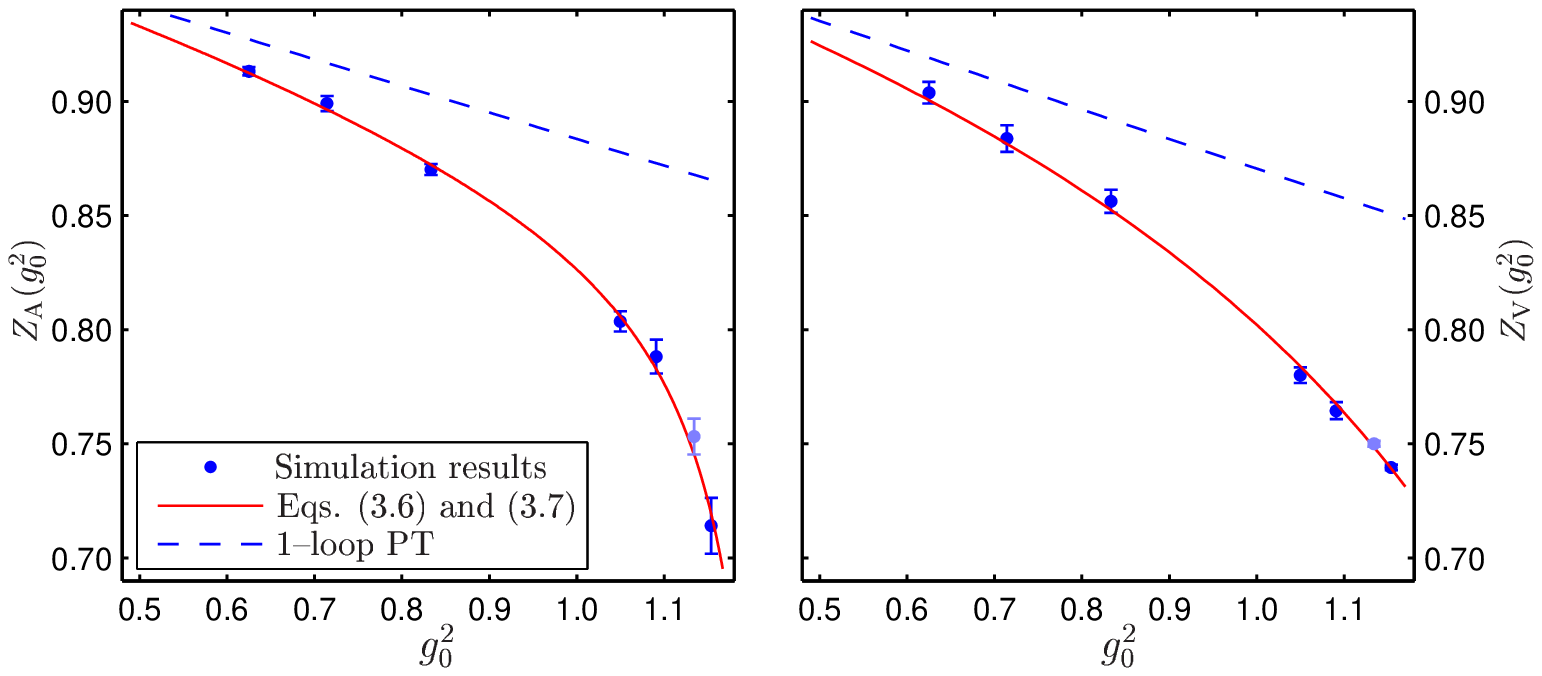,width=150mm}{
Results for $\za$ and $\zv$ from numerical
simulations and 1--loop perturbation theory
(dashed lines). The Pade fits (solid lines)
are given by (\ref{zaint}) and (\ref{zvint}).
\label{fig:result}}
Our final results are shown in \fig{fig:result} as a function
of $g_0^2$.
The errors plotted there are obtained by linearly
summing the
statistical and systematic errors.
One can see that our data for both $\za$ and $\zv$
lie on smooth curves. These can be parameterized
by the interpolating formulae
\bea
\za(g_0^2)&=&\frac{1 -0.918\,g_0^2+0.062\,g_0^4+0.020\,g_0^6}
{1 -0.8015\,g_0^2}\;,\label{zaint}\\[2mm]
\zv(g_0^2)&=&\frac{1 -0.6715\,g_0^2+0.0388\,g_0^4}
{1 -0.5421\,g_0^2}\;.\label{zvint}\\[-1mm]\nonumber
\eea
To $\zv$ we ascribe an absolute error of 0.005, whereas
for $\za$ the absolute error decreases from
0.01 at $\beta=5.2$ to 0.005 at $\beta=5.7$.
The expressions in (\ref{zaint}, \ref{zvint})
have been obtained by a Pade fit,
constrained by 1--loop perturbation theory (\ref{pta}, \ref{ptv}).
Though slightly mismatched and hence excluded from the fit,
also the $\beta=5.29$ data (lighter color in Figs. \ref{fig:result}
and \ref{fig:asq})
are reproduced
by the interpolating formulae in both cases.
For $\zv$ we find agreement at the $1\%$ level
with the results from \cite{Bakeyev:2003ff}, where isospin
charge conservation is imposed in large volume for matrix elements
of the local vector current among nucleon states in the bare coupling
range $\beta\leq5.4$.

We note in passing that similarly to the quenched case
\cite{Luscher:1996jn}
also here the use of mean--field improved perturbation
theory \cite{Lepage:1992xa} for $Z_{\rm A/V}$ improves the
1--loop approximation.
While the difference to our non--perturbative determination
is small for $g_0^2<1$, it rapidly increases in the range
of physically relevant bare couplings.
In particular, at a lattice spacing of roughly $0.1\fm$, corresponding
to $\beta=5.2$, our non--perturbative estimate of $\za$
is almost $20\%$ smaller than the 1--loop value ($10\%$ for boosted
perturbation theory).
In the quenched
case \cite{Luscher:1996jn} this difference was roughly a factor two
smaller at the same lattice spacing.\footnote
{In fact, in \cite{Luscher:1996jn} $\za^{\,\rm con}$ was considered,
but the difference from
our definition of $\za$ was
found to be negligible already for $a$ $\simeq 0.1\fm$.}

\subsection{Systematic effects}

Close to the continuum the dependence of the normalization
factors on the lattice size is expected to be of order $(a/L)^2$
\cite{Luscher:1996jn} in the improved theory.
This implies that effects in $\za$ and $\zv$ due to deviations
from the line of constant physics should be strongly suppressed.

To check for these effects, at $\beta=5.5$ and $5.715$ the
simulations closest to the critical point were
repeated on smaller lattices
($L/a=8$ at $\beta=5.5$ and
$L/a=12$ at $\beta=5.715$) in order
to numerically assess the derivative of $Z_{\rm A/V}$ with
respect to $L$.
Moreover, we estimate the uncertainty in $L$ (measured in units of
$L$ at $\beta\!=\!5.2$) due to
our approximate matching to increase linearly in $\beta$ up to
at most $10\%$ in the range $5.2<\beta<5.715$
(see \cite{DellaMorte:2005se} for a discussion of this
estimate). We  therefore
assign a $6\%$ error to $L$ at $\beta=5.5$ and $10\%$ at
$\beta=5.715$. Together,
this gives the systematic errors quoted in
\tab{t_simpar} through a linear propagation of the error.
Our results confirm the expected small volume dependence.

For the runs with $\beta\geq7.2$ the matched $L/a$ would be
extremely large. On the other hand the volume dependence should
decrease like $g_0^2\,(a/L)^2$ as we are approaching the perturbative
regime.
In practice the simulations are thus performed at $L/a=8$ and
the systematic error is
estimated from additional runs at the coarsest of these lattice spacings,
i.e. at $\beta=7.2$, by taking the difference of $\za$ and $\zv$ between
$L/a=8$ and $16$. While for $\za$ the volume dependence
is hardly visible, in the case of $\zv$ the larger volume
($L/a=16$) results in a statistically significantly lower value.
However, even
this amounts only to a $0.5\%$ effect in the final result.
In addition, we checked the dependence of $\zv$ on the background
field (BF), observing only a moderate effect when going from
BF$=$0 to BF$=$A (used to define the running coupling $\bar g^2$
in\cite{DellaMorte:2004bc}) even for $5.2\leq\beta\leq5.4$.

\subsection{Comparison with an alternative normalization condition}

As discussed in \sect{sectnew} the disconnected diagrams in the
correlation functions (\ref{fXY}) are expected to contribute
only $\rmO(a^2)$ to $\za$.
Hence they can be dropped to obtain an alternative
normalization condition for the axial current.

\EPSFIGURE[!h]{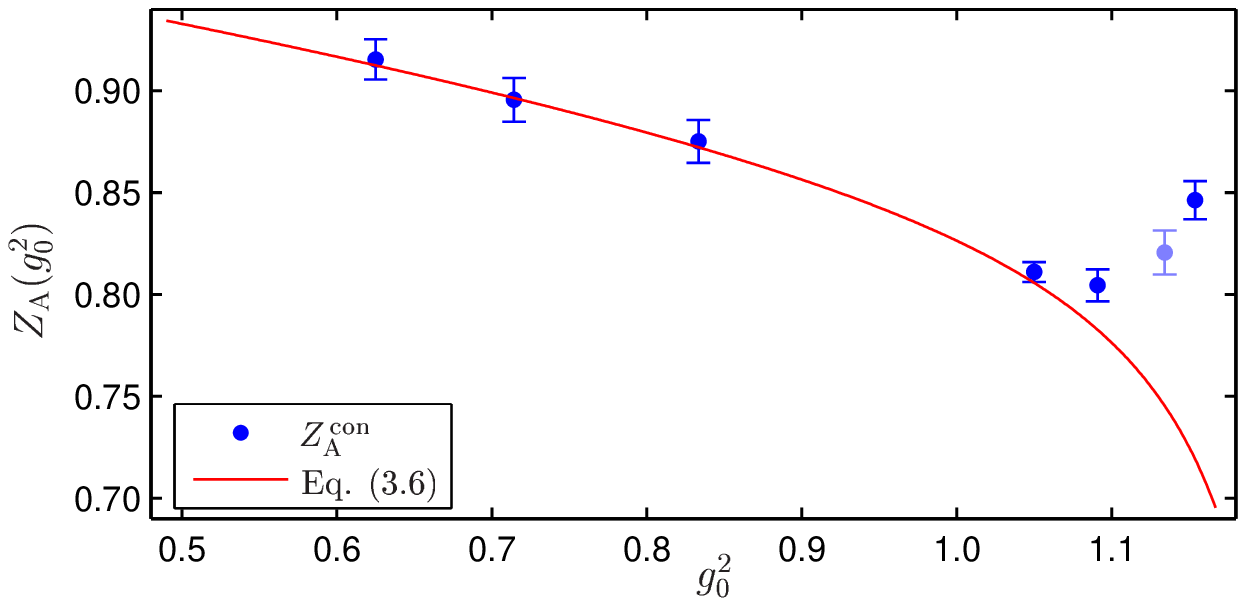,width=110mm}{
Comparison of $\za^{\,\rm con}$ and the interpolating formula
(\ref{zaint}) for $\za$.
\label{fig:asq}}

We carry out the previously discussed analysis also with
this definition to obtain the results shown in \fig{fig:asq},
where they are compared to the interpolating formula
(\ref{zaint}).
The large difference for $g_0^2\gtrsim1.1$ has to be
interpreted as a cutoff effect and we could
indeed explicitly verify that it vanishes
\emph{faster} than linear in $a$.
While this is consistent with the expectations from an improved
theory, the magnitude of these lattice artifacts
is still worrisome.
These findings add further evidence
that for $a\simeq0.1\fm$
cutoff effects with dynamical (improved) Wilson fermions
can be unexpectedly large \cite{Sommer:2003ne}.

%% file: sect4.tex
\section{Conclusions and outlook}
\label{conclusions}
In this work we have shown that in a lattice theory
with two flavors of Wilson fermions normalization
conditions can be imposed at the non--perturbative level
such that isovector chiral symmetries are realized
in the continuum limit.
Since we are working with an improved theory, chiral
Ward--Takahashi identities are then satisfied up to
$\rmO(a^2)$ at finite lattice spacing.

The normalization condition was implemented in terms
of correlation functions in the Schr\"odinger functional
framework and evaluated
on a line of constant physics in order to achieve
a smooth disappearance of the $\rmO(a^2)$ uncertainties.
Through additional simulations at
very small lattice spacings and volumes we verified
that our non--perturbative definition approaches
the perturbative prediction at
small bare gauge coupling.

Simulations were done at or near the critical point
and owing to the new normalization condition, which
keeps track of the mass term in the PCAC relation, any
chiral extrapolations are extremely flat.
Systematic effects due to deviations
from the constant volume condition are
also estimated and turn out to be small.
The results are well described by an interpolating
formula $\za(g_0^2)$ in the range of bare couplings
considered.
Enforcing isospin symmetry with the same programme,
we obtain at the same time a non--perturbative determination of
the normalization constant $\zv(g_0^2)$ of the vector current.
Within about $1\%$ it is in agreement with $\zv$ determined
in \cite{Bakeyev:2003ff} but extends to weak couplings, where contact with
perturbation theory is made.

We found rather large $\rmO(a^2)$ uncertainties in $\za$ at
$\beta<5.4$ by varying the definition
of $\za$.
Together with the algorithmic issues discussed in
\cite{DellaMorte:2004hs} these findings corroborate the worries
expressed in \cite{Sommer:2003ne} about the
status of simulations with improved dynamical Wilson fermions at the
currently accessible lattice spacings.
This merely emphasizes that cutoff effects in physical
quantities can be controlled only if a continuum
extrapolation with several lattice resolutions is
performed.

The result obtained here is an essential step in the
computation of the pseudo--scalar meson decay constant
$F_{\rm PS}$ needed to reliably convert the $\Lambda$ parameter
from \cite{DellaMorte:2004bc} into physical units.
In the short term, together with data from \cite{Knechtli:2002vp}
$\za(g_0^2)$ will be used in a fully
non--perturbative calculation
of the strange quark mass following the strategy
of \cite{Garden:1999fg}.

Finally, the method employed here can also be
used to obtain $\za$ in the $\rmO(a)$--improved
three flavor theory
with either Iwasaki or plaquette gauge action
\cite{Aoki:2002vh,Yamada:2004ja,Ishikawa:2003ri}.

%% file: disco.tex
\section[Contribution of disconnected quark diagrams]
{Contribution of disconnected quark diagrams\footnote{
We thank Martin L\"uscher and Stefan Sint for their contributions
to a clarification of this issue.
}}

\label{appA}

In \fig{fig:contractions} the diagrams, which are
related by an exchange $y_0 \leftrightarrow x_0$ have the
same isospin factors with opposite signs. This follows directly from
(\ref{fAAI}), where such an exchange corresponds
to $\epsilon^{abc}\!\leftrightarrow\epsilon^{bac}$.
Following \cite{disco}, we now argue that in the massless
continuum limit the
contributions of the disconnected diagrams b) and c) to
$\func{AA}(x_0,y_0)$ cancel.
Considering e.g. diagram b), its continuum version is proportional to
\be
f_{\rm AA}^{\,\rm b)}(x_0,y_0)=\int\!\!\rmd^3\bx\,\rmd^3\by
\,\Big\langle \oprime_{ud} (A_0)_{du}(x) (A_0)_{cs}(y)\O_{sc}\Big\rangle\;,
\label{A1}
\ee
where we have introduced flavor indices $u$, $d$, $s$ and $c$ for the
valence quarks, such that diagram b) is the only possible Wick contraction.
Since the spatial insertion points $\bx$ and $\by$ are integrated over,
$f_{\rm AA}^{\,\rm b)}$ depends on $x_0$ and $y_0$ only.
In fact, as we are in the chiral limit and
the operators $\op$ and $\oprime$ generate zero--momentum states,
the axial current is
conserved and hence the diagram is independent of the insertion
points in the two regions $x_0<y_0$ and $x_0>y_0$. If the two
points meet, contact terms may arise, which we need to treat
separately.

To this end we restrict the spatial integration to $|\bx-\by|>\epsilon$
and let $x_0$ approach $y_0$ from either region.
No contact terms can appear due to the finite spatial separation.
In the limit $\epsilon\rightarrow0$ the contribution to
$f_{\rm AA}^{\,\rm b)}$
from the region $|\bx-\by|\leq\epsilon$ vanishes if the integrand
has a divergence weaker than $|\bx-\by|^{-3}$.
In this case we can safely take the limit and conclude that
the order of $x_0$ and $y_0$ does not play any r\^ole.
This would imply that 
the diagrams b) and c) have the same value and since their isospin factors
have opposite signs, their contributions to $\func{AA}$ cancel.

It is clear that the flavor assignment in (\ref{A1}) excludes
a single quark bilinear as
the leading contribution in the short distance expansion
(for $x\rightarrow y$) of
$A_0(x)A_0(y)$. Hence, by power counting, the latter has (if any)
a divergence weaker than $|\bx-\by|^{-3}$ in the limit
$|\bx-\by|\rightarrow0$ and
the contribution from the excluded integration region vanishes.

Since the correlation functions approach their continuum value
with a rate proportional to $a^2$, we can conclude that on the
lattice the contribution from the disconnected diagrams is a
cutoff effect of this order.

%% file: table.tex
\begin{table}[!ht]
\centering{\small
\begin{tabular}{|llrrr@{$\cdot$}lc|llll|}
\hline
\multicolumn{1}{|c}{$\beta$}  & 
\multicolumn{1}{c}{$\kappa$}  &
\multicolumn{1}{c}{$L$}  &
\multicolumn{1}{c}{$T$}  &
\multicolumn{2}{c}{$N_{\rm meas}$}   &
\multicolumn{1}{c|}{$\!\!\!\!\tau_{\rm meas}\!\!$}   &
\multicolumn{1}{c}{$am$}  &
\multicolumn{1}{c}{$\!\!\!\!Z_{\rm A}$}  &
\multicolumn{1}{c}{$\!\!\!\!Z_{\rm A}^{\rm con}$}  &
\multicolumn{1}{c|}{$\!\!\!\!Z_{\rm V}$}
\\\hline
5.200 & 0.13550 & 8 & 18 & 16 & 200 & 4 &$\phantom{-} 0.01718(90)$  & $\!\!\!\!$  0.7301(173)  & $\!\!\!\!$  0.8411(80)  & $\!\!\!\!$  0.7509(6)\\[0mm]
5.200 & 0.13550 & 8 & 18 & 16 & 40 & 10 &$\phantom{-} 0.0159(11)$  & $\!\!\!\!$  0.7186(295)  & $\!\!\!\!$  0.8455(108)  & $\!\!\!\!$  0.7497(14)\\[0mm]
5.200 & 0.13560 & 8 & 18 & 16 & 225 & 3 &$\phantom{-} 0.01310(68)$  & $\!\!\!\!$  0.7157(137)  & $\!\!\!\!$  0.8212(96)  & $\!\!\!\!$  0.7471(7)\\[0mm]
5.200 & 0.13570 & 8 & 18 & 16 & 230 & 2 &$\phantom{-} 0.0088(11)$  & $\!\!\!\!$  0.7134(126)  & $\!\!\!\!$  0.8302(70)  & $\!\!\!\!$  0.7447(8)\\[0mm]
5.200 & 0.13580 & 8 & 18 & 16 & 230 & 2 &$\phantom{-} 0.00194(81)$  & $\!\!\!\!$  0.7176(114)  & $\!\!\!\!$  0.8588(99)  & $\!\!\!\!$  0.7424(14)\\[0.2mm]\hline
5.290 & 0.13625 & 8 & 18 & 16 & 50 & 2 &$\phantom{-} 0.0031(18)$  & $\!\!\!\!$  0.7527(102)  & $\!\!\!\!$  0.8103(167)  & $\!\!\!\!$  0.7507(19)\\[0mm]
5.290 & 0.13641 & 8 & 18 & 16 & 120 & 2 &$-0.00512(61)$  & $\!\!\!\!$  0.7540(124)  & $\!\!\!\!$  0.8378(73)  & $\!\!\!\!$  0.7490(12)\\[0.2mm]\hline
5.500 & 0.13606 & 12 & 27 & 16 & 25 & 6 &$\phantom{-} 0.02254(26)$  & $\!\!\!\!$  0.8417(222)  & $\!\!\!\!$  0.8077(26)  & $\!\!\!\!$  0.7853(14)\\[0mm]
5.500 & 0.13650 & 12 & 27 & 16 & 44 & 3 &$\phantom{-} 0.00758(27)$  & $\!\!\!\!$  0.7987(153)  & $\!\!\!\!$  0.8100(45)  & $\!\!\!\!$  0.7738(8)\\[0mm]
5.500 & 0.13672 & 12 & 27 & 16 & 80 & 3 &$\phantom{-} 0.00041(25)$  & $\!\!\!\!$  0.7888(32)  & $\!\!\!\!$  0.8048(54)  & $\!\!\!\!$  0.7650(21)\\[0mm]
5.500 & 0.13672 & 8 & 18 & 1 & 318 & 4 &$-0.00168(62)$  & $\!\!\!\!$  0.8105(64)  & $\!\!\!\!$  0.8168(38)  & $\!\!\!\!$  0.7750(45)\\[0.2mm]\hline
5.715 & 0.13665 & 16 & 36 & 1 & 106 & 2 &$\phantom{-} 0.00194(57)$  & $\!\!\!\!$  0.8142(135)  & $\!\!\!\!$  0.8079(31)  & $\!\!\!\!$  0.7827(11)\\[0mm]
5.715 & 0.13670 & 16 & 36 & 1 & 54 & 2 &$-0.00060(69)$  & $\!\!\!\!$  0.8004(26)  & $\!\!\!\!$  0.8120(30)  & $\!\!\!\!$  0.7793(20)\\[0mm]
5.715 & 0.13670 & 12 & 27 & 4 & 62 & 2 &$-0.00100(34)$  & $\!\!\!\!$  0.8021(38)  & $\!\!\!\!$  0.8182(18)  & $\!\!\!\!$  0.7861(18)\\[0.2mm]\hline
7.200 & 0.13420 & 8 & 18 & 1 & 220 & 2 &$\phantom{-} 0.00029(45)$  & $\!\!\!\!$  0.8721(24)  & $\!\!\!\!$  0.8772(18)  & $\!\!\!\!$  0.8573(9)\\[0mm]
7.200 & 0.13424 & 8 & 18 & 1 & 164 & 2 &$-0.00028(42)$  & $\!\!\!\!$  0.8683(22)  & $\!\!\!\!$  0.8732(18)  & $\!\!\!\!$  0.8553(6)\\[0mm]
7.200 & 0.13424 & 12 & 27 & 16 & 50 & 2 &$-0.00049(15)$  & $\!\!\!\!$  0.8685(23)  & $\!\!\!\!$  0.8717(8)  & $\!\!\!\!$  0.8543(18)\\[0mm]
7.200 & 0.13424 & 16 & 36 & 1 & 80 & 2 &$-0.00023(41)$  & $\!\!\!\!$  0.8678(18)  & $\!\!\!\!$  0.8670(17)  & $\!\!\!\!$  0.8508(18)\\[0.2mm]\hline
8.400 & 0.13258 & 8 & 18 & 4 & 40 & 2 &$\phantom{-} 0.00023(40)$  & $\!\!\!\!$  0.8990(28)  & $\!\!\!\!$  0.8956(16)  & $\!\!\!\!$  0.8839(15)\\[0mm]
8.400 & 0.13262 & 8 & 18 & 4 & 45 & 2 &$-0.00183(42)$  & $\!\!\!\!$  0.8998(25)  & $\!\!\!\!$  0.8953(13)  & $\!\!\!\!$  0.8826(7)\\[0.2mm]\hline
9.600 & 0.13140 & 8 & 18 & 4 & 100 & 2 &$\phantom{-} 0.00021(15)$  & $\!\!\!\!$  0.9137(14)  & $\!\!\!\!$  0.9154(7)  & $\!\!\!\!$  0.9040(4)\\[0mm]
9.600 & 0.13142 & 8 & 18 & 4 & 125 & 2 &$-0.00059(15)$  & $\!\!\!\!$  0.9118(12)  & $\!\!\!\!$  0.9155(7)  & $\!\!\!\!$  0.9034(4)\\[0.4mm]\hline
\end{tabular}}
\caption{Summary of simulation parameters and results for $\za$ and $\zv$.
\label{longtable}}\end{table}